# Quantitative Measurements of the Influence of Participant Roles during Peer Review Meetings

## Patrick d'Astous, Pierre N. Robillard

École Polytechnique de Montréal, C.P. 6079, Suc. Centre-ville
Montréal, CANADA  H3C 3A7
*dastous@info.polymtl.ca*
*pierre-n.robillard@polymtl.ca*

## Françoise Détienne, Willemien Visser

INRIA-Rocquencour, Bat 24, Domaine de Voluceau,
Rocquencourt, BP 105,
78153, Le Chesnay, France
*Francoise.Detienne@inria.fr*
*Willemien.Visser@inria.fr*

## Abstract

Peer review meetings (PRMs) are formal meetings during which peers systematically analyze artifacts to improve their quality and report on non-conformities.  This paper presents an approach based on protocol analysis for quantifying the influence of participant roles during PRMs.  Three views are used to characterize the seven defined participant roles. The project view defines three roles supervisor, procedure expert and developer.  The meeting view defines two roles: author and reviewer, and the task view defines the roles reflecting direct and indirect interest in the artifact under review.  The analysis, based on log-linear modeling, shows that review activities have different patterns, depending on their focus: form or content.  The influence of each role is analyzed with respect to this focus.  Interpretation of the quantitative data leads to the suggestion that PRMs could be improved by creating three different types of reviews, each of which collects together specific roles: form review, cognitive synchronization review and content review.

**Keywords**:  Peer review meetings, formal technical reviews, process measurement, participant roles, cognitive activities, data analysis, log-linear modeling.

# 1. Introduction

It is helpful if the team activities defined in a software process facilitate the flow of information, which is based on communications, and knowledge, which is based on understanding, among the project team members. Informal and formal communications are two forms of internal team communication that facilitate project progress.

Informal communications, by the definition of "informal," are rarely explicit or prescribed practices of software process, and constitute an implicit activity of any team interaction and a mechanism for maintaining the flow of information and ideas. Typical informal communications are peer-to-peer conversation, electronic mail and informal brainstorming meetings. Little is known about the influence of informal communications in such software development settings, and more studies are needed for a better understanding of their impact on the efficiencies of software development projects.

Formal communication consists of practices often prescribed in software processes in the form of different types of meetings, such as walkthroughs, inspections and review meetings, which we classify under the generic name of peer review meetings or PRMs. The activities expected to take place during a PRM have been outlined in various references (IEEE, 1993) (Bell, 1987).

PRMs are held throughout the development process to verify the content of an artifact resulting from the current phase of the development process, and to validate the specifications for succeeding tasks. By our definition, a project team can hold many PRMs during a week, and each PRM can last anywhere from less than an hour to almost a day.

Using our definition, the number of participants can range from two to the full team. The ideal size of a reviewing team is, however, a subject of debate. Weller (1994) showed that a team composed of four reviewers is twice as efficient as a three-reviewer team. By contrast, Buck (1981) demonstrated that there is no difference in the efficiency of two, three- and four-reviewer teams. Porter *et al.* (1997) concluded that reviewing team size does not influence the anomaly detection rate, and that anomaly detection techniques are the main factor influencing PRM efficiency. Two remaining factors (Porter *et al.*, 1995) must be understood if the efficiency of PRMs is to improve: the cost-benefit ratio, and the factors responsible for the increase in benefits or the reduction in costs have to be identified.

Participants in a PRM may know that roles are influential. It is often observed that some participants are better than others at leading a team meeting. It is also suspected that participants' places in the hierarchy are somehow related to the amount of talking they do. These are both general qualitative observations about what is really going on. To obtain quantitative data on meeting activities finer measurements are needed. Of course, there is an inherent difficulty in setting up an experiment: the participants will be aware of the roles they are playing. Another approach is to observe the PRMsthat are held during a real software development project. In such a case, fewer parameters are under control but the data are from real-life meetings. At the same time, the data from such a case study must be treated with caution, since such a study is a specific set-up. Nevertheless, it could serve to illustrate the benefit of using quantitative data and provide some support for the qualitative interpretation of roles. More and specific case studies are needed, however, before generalized models can be proposed. The analysis presented in this paper is a first step towards quantitative analysis of the PRM.

This paper presents an approach to the study of the PRM based on protocol analysis in which the collaborative activities involved in reviewing a technical document are measured. The basis for this approach is to obtain quantitative data, which enables modeling of the roles and factors that influence the PRM. The working hypothesis is that the influence of roles in a PRM can be quantitatively measured and modeled. Generalization of the model is beyond the scope of this paperbecause the data are based on this particular case study.

## 2. Case-Study Setting

This case study is taken from a real software development project to develop a business process simulator based on Petri Nets. The project required four full-time software engineers and lasted one year. The case study is based on the first nineteen weeks of the project, the period needed to build the beta version of the product. The project team used a defined and documented software engineering process estimated to be at a Software Engineering Institute Capability Maturity

Model process level 2 (Paulk, 1993). An object-oriented paradigm was used throughout the development process. Attendance at PRMs was mandatory and the meetings were required to be held before each artifact composing a milestone was accepted. The results presented in this paper are based on the observation and analysis of seven representative meetings out of the fifteen that were recorded. These seven meetings were chosen both for their commonality - same participants, similar development activities and reviewed solutions, and their differences - the meetings were held at different times in the project lifecycle (two at the start of the project, three in the middle and two at the end).

## 2.1 Peer review process

This study focuses on PRMs held following each design activity of the development process. Acceptance of the artifact at the end of the meeting was a necessary condition of going on to the next step, which is coding. Most of the design activities were performed individually, although some cooperative design was needed from time to time. The resulting design documents were composed of literal descriptions (natural language), algorithms and formal OO notations.

All four team members participated in the observed PRMs. The author of the artifact was to distribute it long enough in advance so that everyone could read it before the start of the PRM. The pace of the meeting was dictated by the time spent on the various sections of the document. A meeting ended on a decision as to the general acceptability of the document.

## 2.2 Definition of participant roles.

A PRM's main objective is to facilitate the peer validation of a document. It is assumed that participant interplays exist and that one individual's action influences another's. Each individual plays a role during the meeting. Table 1 summarizes and compares various meeting roles found in the literature. A document producer, whose role is well recognized, is the author of the document. In collaborative work, the producer represents the team. The role of the moderator is to manage, or lead, the meeting. It is the secretary's role to record the minutes of the meeting, the anomalies found and the action to be taken. The main role of the participants is to review the document, and so any of them can assume the role of reviewer. This role may also be referred to as inspector or specialist.

Table 1 Meeting role definitions in the literature

| Structured Walkthroughs (Yourdon, 1989) | Code Inspection (Fagan, 1976) | Active Design Review (Parnas and Weiss, 1987) | Inspection (Humphrey, 1989) |
|---|---|---|---|
| Producer | Producer Moderator Reader Secretary | Producer | Producer Moderator |
| Reviewer | Inspector | Specialist | Reviewer |

These role definitions consider only the viewpoint of the reviewing process. However, interactions during a meeting involve roles that are played at various levels within the organization and the project. In a small team environment, the same individual can play many roles, depending on the viewpoint. Some roles do not depend on the activity currently being performed by the individual; for example, the role of project manager is the same for the entire project. Other roles are activity-sensitive and are defined by the current one; for example, an individual is assigned the role of moderator for a given meeting. In other cases, the role depends on the subject being discussed. For example, when real estate agents are talking about a house, their roles are likely to be different than if they are talking sellers or buyers. The roles in this case study are defined in Table 2. (d'Astous, 1999).

**Table 2 Definition of roles.**

| Level | Role | Abr. | Definition |
|---|---|---|---|
| **Project** | Procedure Expert | Exp | A designated team member in charge of the application of standards and guidelines for the project. |
| | Developer | Dev | Team members who do not have any specific project responsibilities other than their individual work. |
| | Project Supervisor | Sup | A designated team member who, as part of his work, supervises other team members' technical work. He has a global understanding of the project. |
| **Meet** | Author | Aut | The team member who authored the document being reviewed. |
| | Reviewer | Rev | Team members who are not the author of the document. |
| **Task** | Directly involved | Dir | A participant will have a direct interest in the reviewed artifact if its content is closely related to his task. |
| | Indirectly involved | Ind | A participant will have an indirect interest in the reviewed artifact if its content is not closely related to his task. |

Project roles are always assumed by the same individuals throughout all PRMs. For all meetings analyzed in this study, the project supervisor, the procedure expert and the same two developers were present. The role of project supervisor was more that of a technical overseer than that of a managerial supervisor. The supervisor's responsibilities required a comprehensive and general knowledge of the product being developed. Meeting and task roles changed according to the artifacts under review. Task roles depended on the relationship of the content of the reviewed artifact with the content of the artifacts that the developer, who is acting as reviewer, is responsible for in his development tasks.

Every meeting participant is described in terms of all three roles (project, meeting, task). The testable hypothesis is that these roles have a quantitative and measurable impact on the way in which a meeting proceeds.

## 3. Research Approach

The methods used in this analysis are both quantitative and qualitative. The three steps are necessary to enable quantitative interpretation of the results: data collection, data representation and data analysis,. Raw data are collected and represented in a formal vocabulary (qualitative) and then analyzed using statistical tools (quantitative).

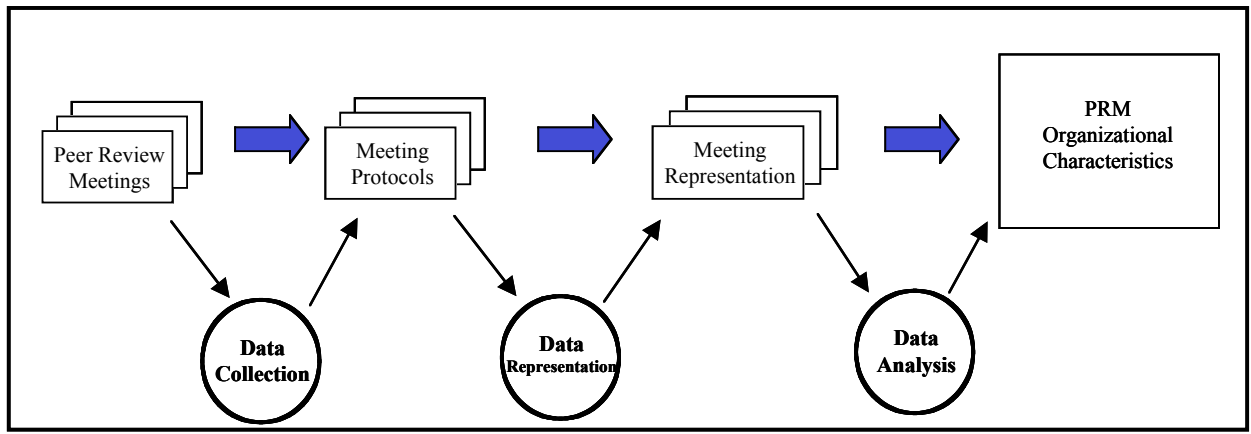

**Figure 1 Main steps in the empirical approach**

The approach is to observe the activities of team members during a PRM and to use the data to derive quantitative measurements or to build a model of the team's behavior during the meetings. This approach is called "Exploratory Sequential Data Analysis" (ESDA) (Sanderson 1994). It is applicable to the analysis of systems, environments or behavioral data whose sequential integrity has been preserved. Figure 1 presents a global view of the research approach. The boxes represent the transformational process from the observed meetings to the resulting PRM model.

The observational approach used was to videotape peer review meetings. A specially trained typist then transcribed the videos. Each individual verbal action, which is called a move, is a transcript entry. Moves may be very short and monosyllabic, like "No" or "Yes", or extended and last for a few minutes. These transcripts form the basic data for the analysis, although it occasionally proved useful to refer back to the video to validate the meaning of some sentences. The protocol, which is made up of the transcription, is a written representation of all the moves. Protocols capture the content of the statements rather than the rationale of their occurrence during the meeting. A team of experts composed of psychologists and software engineers has developed the protocol analysis and the coding scheme used to characterize the moves. The details of this experimental approach have been published elsewhere. (D'Astous 2000, Détienne 1999, Robillard 1998A, Robillard 1998B).

In this section the terminology used in this paper is briefly reviewed (Edmonson, 1981). A sequence is a series of successive moves that deals with a common subject. This paper reports on the analysis of 127 sequences, where the sections of the document under review define the subjects.

A move, which is the contribution of a single speaker to a given sequence, is described by four components:

ID/ACTIVITY/SUBJECT/ATTRIBUTE

1. ID:  identifies the move (speaker and rank in the conversation),

2. ACTIVITY:  identifies the action intended by the speaker,

3. SUBJECT:  defines the entity on which the activity (characteristic 2) is performed,

4. ATTRIBUTE:  (optional) complements the entity with respect to the form or the content (characteristic 3).

The number of activities is limited by the very nature of the review meeting (Robillard, 1998A). Table 3 defines a limited set of activities for the review sequence moves.

**Table 3 Types of activity found in review sequence moves**

| Activity | Abr. | Definition |
|---|---|---|
| Evaluation | EVAL | Judging the value of a subject. This evaluation can either be negative, positive or neutral. |
| Justification | JUSTIF | Arguing or explaining the rationale for a certain choice. It is often necessary to follow up an evaluation with a justification of the approach taken. |
| Information | INFO | Providing new knowledge with respect to the nature of a subject. |
| Hypothesis | HYP | Expressing a personal representation of a subject. This representation is made through the use of expressions such as "I believe that…", "I think …" or "…maybe…" |
| Development | DEV | Presenting a new idea in detail. This is considered a creative activity. |
| Introduction | INT | Introducing the subject. |

Attributes are useful in further describing a given move. There are form attributes and content attributes. Form attributes, which are derived from in-house software engineering guidelines, are used when participants discuss the format of an artifact. Content attributes, which are based on standard ISO 9126, are used when participants discuss the technical aspects of the artifact. Figure 2 provides a sample coding of some moves. In the first column is the Identification (ID), which is composed of the speaker identification and the relative time sequence of the move. In the second column is the coded activity, where each move is characterized by one of the activities listed in Table 3. In the third column is the subject, which is the introduction sequence that is linked to a section of the artifact. In the fourth column is the attribute of the subject, which can be either content or form, depending on the coded move. The next three columns are related to each role level of the speaker, as defined in Table 2. The last column is the order of the sequence.

| ID | | ACTIVITY | SUBJECT | ATTRIBUTE | Meeting | Project | Task | SEQUENCE |
|---|---|---|---|---|---|---|---|---|
| C | 2 | EVAL | INT1 | CONTENT | Reviewer | Developer | Direct | 1 |
| B | 4 | INFO | INT1 | CONTENT | Reviewer | Procedure | Direct | |
| C | 6 | JUSTIF | INT5 | CONTENT | Reviewer | Developer | Indirect | 2 |
| C | 7 | JUSTIF | INT5 | CONTENT | Reviewer | Developer | Indirect | |
| C | 8 | JUSTIF | INT5 | CONTENT | Reviewer | Developer | Indirect | |
| M | 9 | JUSTIF | INT5 | CONTENT | Reviewer | Supervisor | Indirect | |
| M | 10 | JUSTIF | INT5 | CONTENT | Reviewer | Supervisor | Indirect | |

**Figure 2 Coding sample**

## 3.1 Data analysis

The goal of the quantitative analysis is to identify and measure relationships among the various roles.  Figure 3 illustrates the relationships among the dependent and independent variables.  The dependent variables, shown in the middle of the figure, are the move activities and the move attributes.  Independent variables, shown in the boxes, are composed of the seven participant roles from each of the three levels: project, meeting and task.

Log-Linear Modeling (LLM), a discrete multivariate statistical method, is capable of testing relationships among a number of variables while holding constant the effect of all the other variables in a multi-variable matrix.  The term "log-linear" derives from the fact that logarithmic transformations are used to represent relationships among variables.  Bishop *et al.* (1975) detail the mechanisms used to derive log-linear equations.  LLM is used to build a model from empirical data.  It is hierarchical, building and verifying the final representation with additive sub-models of a multivariate data table.  Verification of the final model is obtained by comparing the expected frequencies with the observed frequencies.  LLM provides researchers with partial measures of association, thus allowing researchers to discover the underlying contributions among variables in a multivariate setting (Reynolds, 1977).  Many Web sites provide an overview of this approach (see, for example, LLM-WWW).

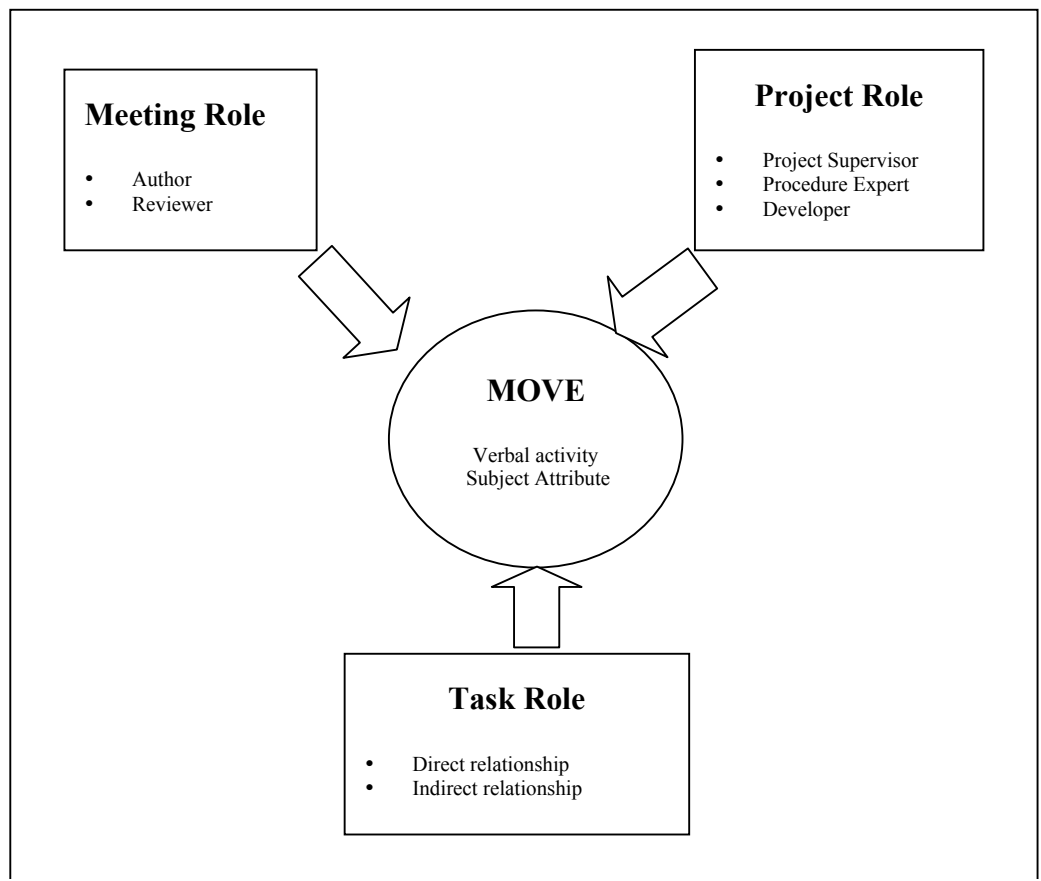

**Figure 3 Influence of independent variables on dependent variables**

## 4. Results

A model built from the analysis of empirical data and validated using LLM is used to identify the factors that influence move occurrences. Results presented in this section were therefore obtained by means of a model derived from empirical values.

### 4.1 Model construction

Model construction requires that all possible relationships among variables (in our case, they are all two- and three-way relationships) be tested for their contribution to the model's fit. Three-way relationship among A B and C implies the three pair-wise relationships among AB, AC and BC.

A model is constructed by adding relationships and measuring its fit with the observed data. A model's fit is measured using the $\chi^2$ statistic. The model used to generate the results was constructed using the three-way relationships identified in Table 4. [But aren't there 24 possible combinations? Isn't, therefore, Table 4 a subset of the possibilities? If so, why?]

**Table 4  Model Relationships**

| DEPENDENT VARIABLE | INDEPENDENT VARIABLE 1 | INDEPENDENT VARIABLE 2 |
|---|---|---|
| Move Activity | Meeting Role | Project Role |
| Move Activity | Meeting Role | Task Role |
| Move Activity | Project Role | Task Role |
| Move Attribute | Meeting Role | Project Role |
| Move Attribute | Meeting  Role | Task Role |
| Move Attribute | Project Role | Task Role |

The quality of the model fit is shown in the scatter diagram in Figure 4. This graph shows expected frequencies using a hypothesis of equal likelihood among the eight outcomes with respect to observed frequencies. A perfectly adjusted model would have all its points located on the segment. Finally, $\chi^2$ is computed using the Likelihood Ratio test. The model used in this analysis yielded a value of 26.4, which is 99% significant for a model with 55 degrees of freedom. The number of degrees of freedom is the total number of variables minus one. There are 8 dependent variables, which are the 6 move activities plus the 2 subject attributes, and there are 7 independent variables, which are the 2 meeting and task variables and the 3 project variables.

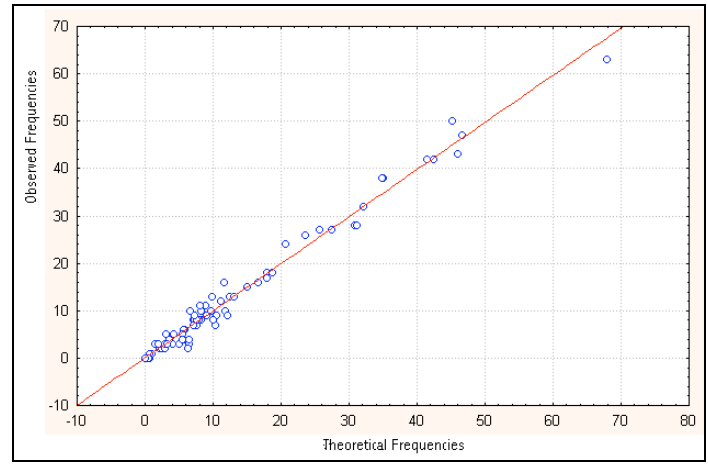

**Figure 4 Scatter diagram of expected or theoretical frequencies vs observed frequencies**

## 4.2    Data analysis

A model such as the one built for this study enables a deeper analysis of phenomena found in PRMs. The meeting roles (author vs reviewer) seem to have an impact on the discussions. Figure 5 shows the significant cleavage created by these roles in the discussions of form and content.

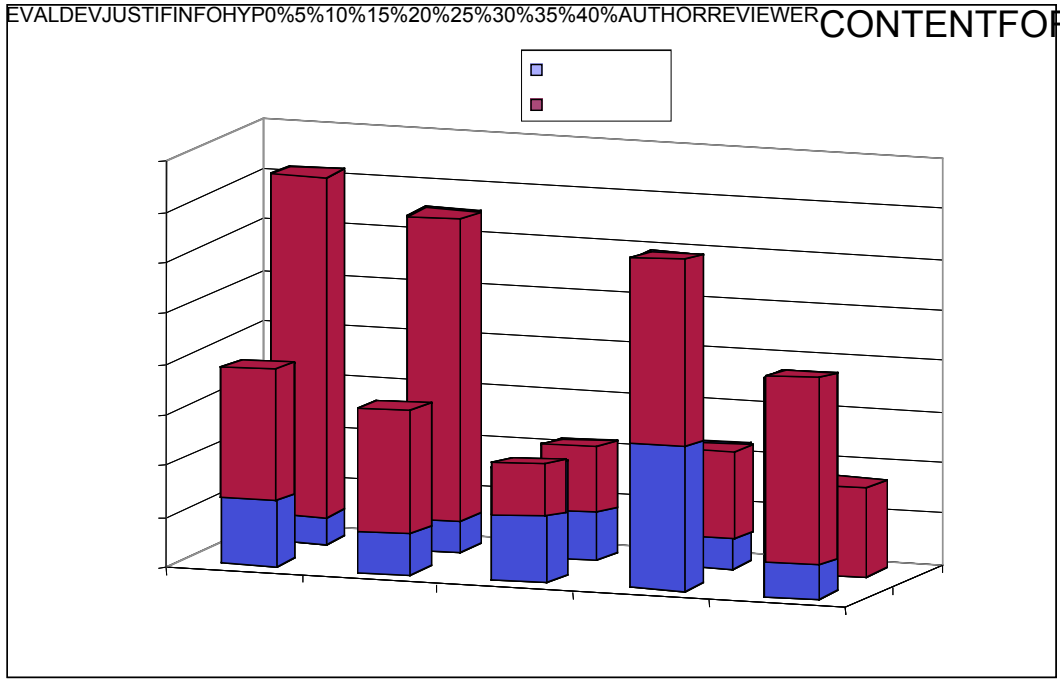

**Figure 5 Relative frequency of occurrence of moves**

The front row represents the relative number of moves related to content for each activity, while the back row represents the relative number of moves related to form. The first column in the first row (left-front column) shows that almost 20% of the moves related to content stem from EVAL activities. The lower part of this column shows that almost 5% of these moves come from the author and the rest from the reviewers. The relative contribution of the author to a form discussion is small (only 15%), as shown by the sum of the lower part of the back row columns, while the author's contribution climbs to 35% when the discussion is related to content.

A detailed analysis of the move activity can highlight certain behaviors. Moves for the justification activity show three unique patterns. First, the total relative number of justification moves is around 10%, whether for a form or a content attribute; second, it is the only activity wherein the author and reviewers have an equivalent (50%) number of moves; and third, this ratio is similar, whether for a form or a content attribute.

It is observed that the author is involved very little in form move activities (back row). It is also observed that almost 70% of all form moves concern evaluation and development activities (Figure 5), while only 17% of form moves serve to increase comprehension (information and hypothesis). Most of the discussions related to the form of the document are corrections (DEV, EVA) proposed by reviewers. These corrections generate little discussion on the part of the author, which may indicate that most of the time they are trivial non-conformities.

Participation patterns are quite different when the team discusses document content. More than 50% of the content move activities are oriented towards understanding of the artifact better; these are the information and hypothesis moves. The author and the reviewers seem to spend more time trying to agree on the content of the document than in proposing corrections (evaluation and development). The author is mostly involved with the information moves, which is what is expected when someone is explaining his or her work.

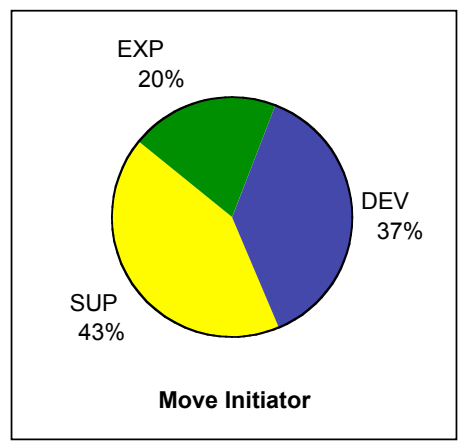

**Figure 6 Relative distribution of move initiation**

Figure 6 shows a pie-chart distribution of the roles that are responsible for initiating move activities. It is observed that the project supervisor (SUP) makes most of the moves (43%) related to the document under review. The two developers together (DEV) account for only 37% of the moves. This result is consistent with the findings of McGrath (1984), in which he states that during small group meetings the most frequent initiator of conversations will initiate 40% to 45% of all moves. Herbsleb *et al.* (1994), in their study of object-oriented design meetings, also mention that the participation pattern of the *chief architect* (roughly equivalent to the project supervisor) is consistent with McGrath's findings. Thus, the project supervisor's role in the document review process may have quite a strong influence. The personality of the participants was not considered in this study.

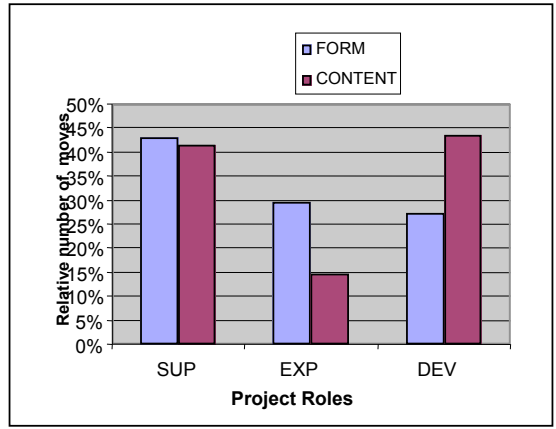

**Figure 7 Move initiation vs move attribute**

Figure 7 presents project moves relative to the attributes of form and content. The first pair of columns on the left shows that the participation of the project supervisor (SUP) in discussions of form is relatively similar to that in discussions of content. The procedure expert is twice as involved in the form move as in the content move. He initiated 30% of the form moves, as compared to only 15% of the content moves. Greater involvement in the form attributes is indeed expected from the procedure expert. Developers are relatively more involved in the content of the document under review than in its form.

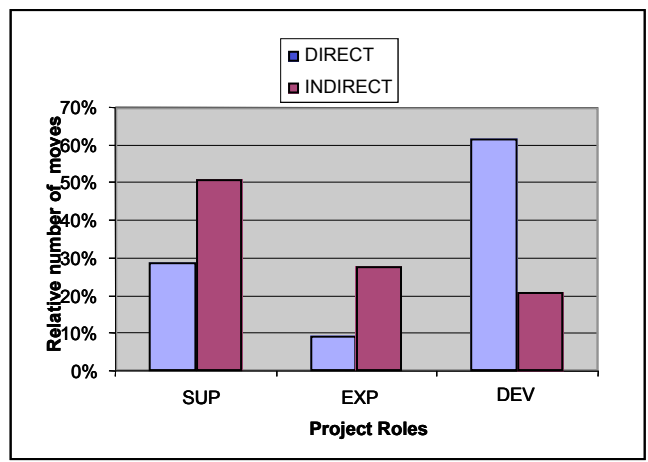

**Figure 8 Move initiation vs project task**

Figure 8 presents the moves associated with the project roles as they relate to the task roles. The purpose is to see if the task roles have any influence on the project roles. The first column shows that the supervisor is responsible for almost 30% of all the moves concerning an artifact with which he is directly involved. In this project, the supervisor was also responsible for developing some software. It appears that the project supervisor and the procedure expert participate relatively more in discussions related to a documentthat has an indirect relationship with their own individual tasks. This behavior is expected since the same individuals assume these two roles for the project duration. However, it is interesting to observe that developers are three time less active when the artifacts are not related directly to their own work.

## 5. Discussion

Most technical review meeting approaches define basic roles, such as the author and the reviewers. Fagan (1976) went further by defining more meeting responsibilities (reader, moderator, etc.) for participants. Results in this paper show that there are other factors that may influence the progress of technical review meetings.

Participant involvement in a PRM seems to be influenced by the roles played within the development team. This case study outlines two activities (or meta-activities) for the review artifact. The first activity, the form review, is expeditious and characterized by the evaluation and development of simple alternatives that do not require lengthy discussion. According to our study form reviews are implicitly accepted most of the time by the author of the document. The second activity, content review, is for the most part a cognitive synchronization (Falzon, 1996) activity (information and hypothesis), wherein participants adjust their own views to those of others. Document content is seldom reviewed without prior consensus on the meaning of this content.

Form and content are not reviewed in the same way. Form review is quick, and performed as stipulated in the literature, whereas for content review a PRM would seem to be an opportunity for all participants to synchronize their views and move towards the common goal. While both form review and content review are necessary, they may not have to be performed during the same meeting. Form review could be conducted outside the PRM itself by a single individual (procedure expert), for example, who would submit changes to the author. Once form review has been completed, the team could then meet to revise the technical content of the document.

Review, as a cooperative decision-making activity (Fisher, 1974), is highly influenced by the project supervisor whose responsibilities and global knowledge of the project can more easily bring others over to his or her point of view. The project supervisor's participation in all PRMs is therefore obviously needed. The participation of others may depend on their roles. The presence of the author of the document being reviewed is necessary to improve cognitive synchronization, as demonstrated by his major involvement in providing information. The participation of other team members as reviewers may depend on the relationship of the artifact under review with their own work. Developers acting as reviewers are far less active when reviewing documents that are not directly related to their own tasks. The procedure expert, by the nature of his responsibilities, is mostly concerned with the form of the document. It is found that most activities are provided by the participant who has a defined role (supervisor or expert) or a direct interest in the artifact being reviewed, either as author or as interested user of the information described in the artifact.

Following the idea of Seaman and Basili (1998), it may be claimed that the organizational structure itself influences the nature of communications occurring in PRMs. Our results show that three views seem to influence these communications:

1. Meeting view: The literature describes two principal roles which are active during a PRM: those of author and reviewer. This study shows quantitatively that other roles, such as those of supervisor and procedure expert, may have a major impact on meeting outcomes.

2. Project view: This study shows that individual project responsibilities may influence participant moves. An efficient PRM should involve the appropriate participants, and these participants should be selected according to their project roles.

3. Task view: The suitable number of participants taking part in a PRM is still being debated in the literature. This study shows that it is less a matter of the number of participants than of selecting the appropriate participants. Participants with a direct interest in the artifact being reviewed are the ones who will be the most active during the review process. This direct interest could be evaluated by the level of the relationship that exists between their own work or task and the artifact being reviewed.

## 6. Conclusion

The PRM is becoming a key practice in the software development process, and PRM efficiency could be easily measured by recording the anomaly detection rate. However, this study shows that the efficiency of a PRM is also affected by the roles of the participants. In fact, quantitative data show that the detection of anomalies related to form is a minor activity, which does not require a PRM, but could be performed by a single procedural expert. A major part of the PRM is spent in providing information and formulating a hypothesis, that is, cognitive synchronization.

A PRM should have at least two participants, who are the author of the artifact and the project supervisor. These two participants represent the minimum resources needed for a PRM. The author provides a detailed understanding of the artifact, and the supervisor a comprehensive view of the project and an understanding of how the artifact content fits into the project. All other team members who have a direct relationship with the document under review may complement this duo. There are advantages to having other team members who are directly impacted attend the meeting: they are likely to contribute alternative solutions, and, more important, all the participants will improve their understanding of the project by synchronizing their views. While increasing the number of participants is likely to improve the quality of the review, it will also increase the cost in terms of the resources participating in the meeting.

The quantitative data provided by this case study provide some insight into the activities carried on during a PRM. In view of the results presented in this paper, a three-level structure could be proposed for PRM meeting. At level one would be form review, which is performed by the procedure expert alone. (The author does not need to participate in the form review, since our data show that this individual reacts very little to comments about form.)

At second review level would be characterized by cognitive synchronization. At this level, every team member interested in the artifact will participate in the meeting. The author and the project supervisor will conduct a walkthrough of the artifact to provide information and to discuss hypotheses related to the content of the artifact.

The third and last meeting would be a defect detection meeting. Ideally, there would be no form discussion and very little cognitive synchronization, since all participants will have attended the level-two meeting. This meeting could take place with a much smaller number of participants than required at the previous level. At a minimum, this meeting would include the author and the supervisor.

## 7. Acknowledgements

We thank the team members who participated in the project and offered their enthusiastic collaboration. We also thank DMR Consulting Group for providing us with the opportunity to gather meaningful data for the project. We are grateful to the lead editor Stan Rifkin for his help in improving the manuscript and the anonymous reviewers for their very appropriate comments.

This project was supported in part by the Canadian STEP program, le programme d'échange France-Québec that enables collaboration with INRIA, NSERC grant A0141 and an NSERC industrial scholarship.

Second International Conference on Design of Cooperative Systems. Juan-les-Pins, France.